\documentclass[a4paper,10pt]{article}

\usepackage{calrsfs,times}
\usepackage{a4wide}
\usepackage{graphicx,psfrag,enumerate}
\usepackage{amssymb}
\usepackage{amsmath}
\usepackage{calrsfs}
\usepackage{color, epsfig, epstopdf, wrapfig}
\usepackage{paralist}
\usepackage{float,cuted}

\newtheorem{theorem}{Theorem}

\newtheorem{lemma}{Lemma}

\usepackage{amsmath}
\usepackage{accents}
\newlength{\dhatheight}

\usepackage{color}
\usepackage[normalem]{ulem}
\usepackage{soul} 
\soulregister\cite7
\soulregister\ref7
\soulregister\pageref7

\begin{document}

\title{\LARGE \bf 
  Resilient Distributed $H_\infty$ Estimation via 
  Dynamic Rejection of Biasing Attacks    
  \thanks{This work was supported by the Australian
  Research Council and the University of New South Wales. The paper will
  appear in the Proceedings of the 2018 American Control Conference.}}

\author{V.~Ugrinovskii\thanks{V. Ugrinovskii is with the School of Engineering and Information Technology, University of New South Wales at the Australian Defence Force Academy, Canberra, ACT 2600, Australia. {\tt\small v.ougrinovski@adfa.edu.au}}}

\maketitle
         
\begin{abstract}
We consider the distributed $H_\infty$ estimation problem with additional
requirement of resilience to biasing attacks. An attack scenario
is considered where an adversary misappropriates some of the observer nodes
and injects biasing signals into observer dynamics. Using a dynamic
modelling of biasing attack inputs, a novel distributed state estimation
algorithm is proposed which involves feedback from a network of attack detection
filters. We show that 
each observer in the network can be computed in real time and in
a decentralized fashion. When these controlled observers are
interconnected to form a network, they are shown to 
cooperatively produce an unbiased estimate the plant, despite some of the
nodes are compromised. 
\end{abstract}

\section{Introduction}

Problems of resilient control and estimation came into prominence
after situations were discovered where an adversary was able to interfere
with the control task by covertly injecting false information into the measurement
data~\cite{MKBDLPS-2012,DS-2010a,
  PDB-2013,ZBB-2013}. 
Networked control systems are particularly vulnerable to data injection
attacks, since operation of such systems 
depends on the integrity of
communicated data. 

The information shared between the nodes can also
be utilized for monitoring integrity of the 
network. We demonstrate in
this paper that the information routinely
collected and shared within a distributed observer network can be used to
detect compromised observers and neutralize their biasing effect. 
For this, we propose a novel distributed observer augmented with a
network of 
attack detector filters; the latter filters
provide feedback to the node observers which neutralizes rogue biasing
inputs, if such inputs are present.  

The model of misappropriation attack considered here is the same as
in~\cite{DUSL1a}; it captures essential features of the biasing attack
described in~\cite{TSSJ-2015}. 
It assumes that
the adversary gains access to one or several nodes of the observer network
and injects biasing inputs directly into the state estimation algorithm; also,
cf.~\cite{TSSJ-2014}. Similarly to~\cite{DUSL1a}, our approach to detecting
such biasing behaviour is by tracking changes in the behaviour of the estimation
errors caused by malicious biasing inputs. However different
from~\cite{DUSL1a}, our method allows each observer node to compute its
attack detection filter in decentralized fashion, without communicating
with other nodes. 

The decentralization of computations is accomplished in this paper using a decoupling
technique which we have developed previously for the 
distributed observer design in~\cite{ZU1a}. The technique involves an 
initial `setup' step which requires the network to compute certain
auxiliary parameters that are then distributed among the nodes.
Although this initial 
setup must be carried out centrally, it involves
only the information about the communication network, and does not require
knowledge of the plant observed. The auxiliary parameters computed at this
setup step are then used for computing a collection of controlled node
observers equipped with output feedback $H_\infty$ controllers; the latter
can detect and cancel the attack using the 
same sensory data that are available for  estimation of the state of the
observed 
plant. 
Also, 
our method 
allows for the consideration of time-varying
distributed filters. The robustness against
uncertainties in the sensors and the plant model 
is guaranteed as
well. 

The feedback-controlled nature of the proposed distributed observer
distinguishes this paper from the companion paper~\cite{U10a}, where we
used a similar decoupling technique.   
Here, the observers are
computed \emph{jointly} with the attack detection filters. This requirement
of co-design did not arise in the attack detection problem considered
in~\cite{DUSL1a,U10a}, but arises in this paper since here the objective
shifts from detecting and signalling a biasing attack to ensuring the
distributed observer network is resilient to this kind of attacks. 

  

\emph{Notation}: $\mathbf{R}^n$ denotes the real Euclidean $n$-dimensional vector space, with the norm  $\|x\|=(x'x)^{1/2}$; here the symbol $'$ denotes the transpose of a matrix or a vector.
The symbol $I$ denotes the identity matrix. For real symmetric
$n\times n$ matrices $X$ and $Y$, $Y>X$ (respectively, $Y\geq X$) means the
matrix $Y-X$ is positive definite (respectively, positive semidefinite). 
The notation $L_2[0, \infty)$ refers to the Lebesgue space of
$\mathbf{R}^n$-valued vector-functions $z(.)$, defined on the time interval
$[0, \infty)$, with the norm $\|z\|_2\triangleq\left(\int_0^\infty
  \|z(t)\|^2 dt \right)^{1/2}$ and the inner product $\int_0^\infty z_1'(t)
z_2(t) dt$. 
   
\section{Biasing misappropriation attacks on distributed observers}
\label{sec:distributed_estimation}

A distributed observer problem consists in
obtaining an estimate of the 
state of a time varying plant 
\begin{eqnarray}
\label{state}\label{eq:plant}
 \dot{x} = A(t)x +B(t)w, \quad x(0)=x_0,
\end{eqnarray}
which is subject to an unknown modeling disturbance $w$. The estimate is to
be obtained from a collection of measurements 
\begin{equation}\label{measurement}
y_i = C_i(t)x + D_i(t)v_i, \quad i=1,2,\ldots,N,
\end{equation}
taken at $N$ nodes of a sensor network, each perturbed by a measurement
disturbance $v_i$. In the distributed estimation setting a state estimate
must be obtained at each 
network node without sending the data to a central data processing
facility, and the nodes must obtain the same estimate of the
plant. This is achieved by interconnecting the 
observers into a network. This way, the nodes can use the information which
they receive 
from their neighbours to correct their plant state
estimates until all nodes reach an agreement. 

Let
the state $x$ and the disturbance $w$ be vectors in $\mathbb{R}^{n}$,
$\mathbb{R}^{m}$ respectively, and each measurement $y_i$ be a vector in
$\mathbb{R}^{p_i}$. The 
disturbances $w$ and $v_i\in\mathbb{R}^{m_i}$ will be assumed to be 
$\mathcal{L}_2$ integrable signals defined on the interval
$[0,\infty)$. The initial state $x_0$ is also assumed to be 
unknown. A typical distributed estimation problem involves a
network of filters connected over a graph with vertices in the set
$\{1,\ldots,N\}$, each representing a node of the
network:
\begin{eqnarray}
\dot{\hat{x}}_i &=& A(t)\hat x_i + L_i(t)(y_i-C_i(t)\hat x_i) \nonumber \\
&&+\sum_{j\in\mathbf{N}_i}K_{ij}(t)(c_{ij}-W_{ij}\hat x_i), \quad
\hat{x}_i(0)=\xi_i. 
\label{filter_i}\label{UP7.C.d.unbiased}
  \end{eqnarray}
Each observer (\ref{UP7.C.d.unbiased}) produces an estimate $\hat x_i(t)$
of the plant state $x(t)$. For this, it uses its measurement $y_i$ and the
information received from the neighbours; the latter information is
communicated over noisy communication channels in the form of 
$p_{ij}$-dimensional signals  
\begin{equation}\label{communication}
c_{ij} = W_{ij}\hat{x}_j + H_{ij}v_{ij},\quad j\in\mathbf{N}_i.
\end{equation}
Since the plant is time-varying, 
the
observer gains  $L_i$, $K_{ij}$ in (\ref{UP7.C.d.unbiased})  are allowed to be time-varying. 

The signals $c_{ij}$ complement the local measurements $y_i$ at node
$i$ and assist it in obtaining a high fidelity estimate of the
plant. Each such signal contains information about the neighbour's estimate
$\hat x_j$ of the plant state $x$. That is, the observers
(\ref{UP7.C.d.unbiased}) are coupled via the signals $c_{ij}$, forming a
distributed observer network. Such a coupling between the observer
nodes is essential in situations where the plant is not detectable from local
measurements at some of the nodes, and these nodes require additional
information which can only be obtained from
their neighbours. 
The matrix $W_{ij}$ determines the part of the vector
$\hat x_j$ which node $j$ shares with node $i$. Since this information is 
usually delivered over noisy communication links, a
disturbance $v_{ij}$ is included in (\ref{communication}) which is also
assumed to be an $\mathcal{L}_2$ integrable signal. 

The task of distributed estimation using the observer network
(\ref{UP7.C.d.unbiased}) is to ensure that each estimate
$\hat x_i(t)$ converges to $x(t)$ as $t\to \infty$ in
some sense, with some robustness against disturbances in the plant
model, measurements and interconnection channels. A large body of literature
is dedicated to the question as to how the observers 
\eqref{UP7.C.d.unbiased} can be constructed which achieve
this objective; e.g.~\cite{Olfati-Saber-2007,DPB-2013,MOVRDJ-2013,MK-2013}.  
However, 
the dependency on information sharing
leaves the distributed observers vulnerable to attacks seeking to disrupt the
estimation task. A scenario of such attacks usually  
considers an injection of false signals into sensor
measurements or communicated data~\cite{PDB-2013}. Here we
follow~\cite{DUSL1a} and consider a different scenario where the adversary
substitutes one or several observers (\ref{UP7.C.d.unbiased})
with their biased versions 
\begin{eqnarray}  
    \dot{\hat x}_i&=&A(t)\hat x_i + L_i(t)(y_i(t)-C_i(t)\hat x_i) \nonumber \\
&& +\sum_{j\in
      \mathbf{N}_i}K_{ij}(t)(c_{ij}-W_{ij}\hat x_i)+F_if_i, \quad \hat
    x_i(0)=\xi_i,\quad 
  \label{UP7.C.d}
\end{eqnarray}
Here $F_i\in\mathbf{R}^{n\times n_{f_i}}$ is a constant matrix and $f_i\in \mathbf{R}^{n_{f_i}}$ is an unknown signal representing the attack
input. 
\begin{figure}[t]
\psfrag{R}{$f_i$}
\psfrag{nu}{$-\nu_i$}
\psfrag{Y}{$\hat f_i$}
\psfrag{H}{$G_i(s)$}
\psfrag{G}{$\frac{1}{s}$}
\psfrag{+}{$+$}
\psfrag{-}{$-$}
  \centering
  \includegraphics[width=0.45\textwidth]{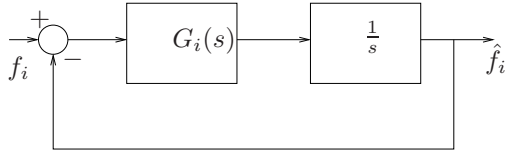}
  \caption{An auxiliary `input tracking' representation of a biasing attack
    input introduced in~\cite{DUSL1a}.} 
  \label{tracker}
\end{figure}
Following~\cite{DUSL1a}, we consider a class of
attacks on the filter (\ref{UP7.C.d}) consisting of biasing
inputs $f_i(t)$ of the form
\begin{equation}
  \label{decomp}
f_i(t)=f_{i1}(t)+f_{i2}(t),  
\end{equation}
where the Laplace transform of $f_{i1}(t)$, $f_{i1}(s)$, is such that
$\sup_{\omega}|\omega f_{i1}(j\omega)|^2<\infty$ and
$f_{i2}\in L_2[0,\infty)$. 
Obviously, biasing inputs with rational Laplace transforms which have no more
than one pole at the origin and the rest of the poles located in the open
left half-plane of the complex plane have this property. We will use the
notation $\mathcal{F}$ for the set of such inputs. It includes
biasing attack inputs introduced in~\cite{TSSJ-2015} 
consisting of a steady-state component and an exponentially vanishing
component generated by a low pass filter.    

The following lemma characterizes the properties of biasing inputs of this
class. Its proof is given in the journal version 
of~\cite{DUSL1a}. Let $G_i(s)$ be a proper transfer function for which the
system in Fig.~\ref{tracker} is stable, and $\hat f_i$ be an output of that
system.     
\begin{lemma}\label{admiss.f}
\begin{enumerate}[(i)]
\item
Consider a class of signals $f_i(t)$ that admit the decomposition
(\ref{decomp}). Then for all such signals $f_i(t)$ it holds that 
\begin{equation}\label{f-eta} 
 \int_0^\infty\|f_i-\hat f_i\|^2dt<\infty.  
\end{equation}
\item
If in addition, $G_i$ is selected so that
\begin{eqnarray}
\lim_{s\to 0}
\|(I+\frac{1}{s}G_i(s))^{-1}\|=0, \label{G} 
\end{eqnarray}
then $\lim_{t\to\infty}\|f_i(t)-\hat f_i(t)\|= 0$ for all inputs $f_i\in
\mathcal{F}$.  
\end{enumerate}
\end{lemma}

According to Lemma~\ref{admiss.f}, biasing inputs that have the form
(\ref{decomp}) can be `tracked' using a system shown in Fig.~\ref{tracker}.
Of course, in reality it is not possible to track covert attack
inputs. Nonetheless, the model in Fig.~\ref{tracker} 
allows us to associate the class of biasing attack inputs with the minimal
realization of the strictly proper transfer function $\frac{1}{s}G_i(s)$,
of the form
\begin{eqnarray}
&&\dot\epsilon_i = \Omega_i\epsilon_i+\Gamma_i \nu_i, \qquad
\epsilon_i(0)=0, \label{Om.sys.general} \\ 
&&\hat f_i= \Upsilon_i\epsilon_i, \nonumber
\end{eqnarray}
where $\nu_i=\hat f_i-f_i$ is an $L_2$-integrable input, according to
(\ref{f-eta}). Clearly, each signal $\nu_i$ corresponds to a certain
unknown biasing input $f_i$; it represents a mismatch error between the
attack input $f_i$ and the output $\hat f_i$ of the system
(\ref{Om.sys.general}). In the sequel, this error will be regarded as an
additional $L_2$-integrable disturbance which will arise when we replace
$f_i$ with $\hat f_i$ in the derivation of our attack detection and resilient
estimation algorithms. 

Apart from ensuring stability of the system in Fig.~\ref{tracker},
according to Lemma~\ref{admiss.f}, the proper
transfer function $G_i(s)$ can be selected arbitrarily. 

\section{Problem Formulation}\label{Problem.formulation}

In this paper we are concerned with the design of resilient version of the
distributed observer (\ref{UP7.C.d.unbiased}). Our approach is to augment
each node observer with additional dynamic feedback controllers to suppress the
attack inputs. 
To accomplish this task, we introduce the following
controlled modification of the observers  
(\ref{UP7.C.d.unbiased}), (\ref{UP7.C.d}): 
\begin{eqnarray}  
    \dot{\hat x}_i&=&A(t)\hat x_i + L_i^r(t)(y_i(t)-C_i(t)\hat x_i) \nonumber \\
&& +\sum_{j\in
      \mathbf{N}_i}K_{ij}^r(t)(c_{ij}-W_{ij}\hat x_i)+F_if_i+u_i,
  \label{UP7.C.d.res} \quad  \\
    \hat x_i(0)&=&\xi_i, \nonumber 
\end{eqnarray}
The superscript $^r$ is to emphasize that the gains $L_i^r(t)$,
$K_{ij}^r(t)$ are to be different from the gains $L_i(t)$,
$K_{ij}(t)$ of the original observer (\ref{UP7.C.d}). Also, $u_i$ denotes   
the control input. 

We propose the following observer-based feedback structure for
generating the controls $u_i$: 
\begin{equation}
u_i=-F_i\varphi_i,
\label{contr}
\end{equation}
Here $\varphi_i$ denotes an output of a filter
\begin{eqnarray}
  \label{detector.general}
  \dot\mu_i&=&\mathcal{A}_d(t)\mu_i + L_{d,i}(t)(\zeta_i-W_{d,i}\mu_i)
  \nonumber \\
  && +\sum_{j\in 
    \mathbf{N}_i}K_{d,ij}(t)(\zeta_{ij}-W_{d,ij}(\mu_j-\mu_i)), \\
  \varphi_i&=&C_{d,i}(t)\mu_i, \qquad
  \mu_i(0)=\mu_{i,0}; \nonumber 
\end{eqnarray}
where $\mathcal{A}_d(t)$, $L_{d,i}(t)$,  $K_{d,ij}(t)$, $W_{d,i}$
$W_{d,ij}$, $C_{d,i}$ are matrix coefficients to be found. Each 
filter (\ref{detector.general}) is governed by the innovation signals 
 $\zeta_i$, $\zeta_{ij}$
:  
\begin{eqnarray}
  \zeta_i&=&y_i-C_i(t)\hat x_i, 
                                              \label{out.y} \\
  \zeta_{ij}&=&c_{ij}-W_{ij}\hat x_i.
                                              \label{out.c}
\end{eqnarray}
The filter (\ref{detector.general}) must
generate $\varphi_i$ so that when node $i$ is under attack, the 
signal $u_i$ counters the biasing input $f_i$. Also, at the nodes which
are not attacked directly, $u_i$ must not interfere with
the state observer. This requires the output $\varphi_i$ of the filter (\ref{detector.general}) to track the biasing signal
 $f_i$, turning (\ref{detector.general}) into  an attack detector.       



The problem of resilient estimation under
consideration is now formally stated as the problem of constructing a
network of filters (\ref{detector.general}) which, when interconnected with 
the modified state observers (\ref{UP7.C.d.res}) via the feedback control
(\ref{contr}), achieve the following properties
\begin{enumerate}[(i)]
\item
In the absence of disturbances and when the system is not under attack, at
every node $i$, $\|x(t)-\hat x_i\|$ and $\varphi_i$ converge to 0
exponentially.  
\item
In the presence of uncertainties and/or attack, 
\begin{eqnarray}
&&\int_0^{+\infty}\|\varphi_i-f_i\|^2dt < +\infty \quad \forall i, \nonumber \\ 
&&\int_0^{+\infty}\mathbf{e}'P\mathbf{e}dt <+\infty;
\label{convergence}
\end{eqnarray}
here $P=P'\ge 0$ is an
$nN\times nN$ matrix and $\mathbf{e}=[e_1'~\ldots~e_N']'$, where
$e_i=x-\hat x_i$ denotes the estimation error of the observer
(\ref{UP7.C.d.res}) at node $i$. 
\end{enumerate}

The first condition in (\ref{convergence}) formalizes the requirement for
the filters 
(\ref{detector.general}) to track the corresponding 
attack inputs in the $\mathcal{L}_2$ sense. Therefore, by monitoring 
the behaviour of the outputs $\varphi_i$, it will be possible
to establish which node has been attacked. The second condition in (\ref{convergence}) describes the desired resilience property of the observers.
The matrix $P$ is considered to be given
. The resilience of the modified observers
(\ref{UP7.C.d.res}), (\ref{contr}), (\ref{detector.general}) requires 
(\ref{convergence}) to hold for any collection of admissible biasing inputs
$f_i$ 
described in Section~\ref{sec:distributed_estimation}.   
The problem in this paper is to determine  the
characteristics  $\mathcal{A}_d(t)$, $L_{d,i}(t)$,  $K_{d,ij}(t)$,
$W_{d,i}$, $W_{d,ij}$, $C_{d,i}(t)$ of the filter (\ref{detector.general}) which
guarantee that the above conditions (i) and (ii) hold.



\section{Design of resilient distributed observers}\label{resilient.estimation}
 
\subsection{Analysis of error dynamics}

To construct suitable filters (\ref{detector.general}) consider the
dynamics of the estimation errors of the controlled observers
(\ref{UP7.C.d.res}), (\ref{contr}). It is easy to see that these errors evolve
according to  
 \begin{eqnarray}
    \dot{e}_i&=&(A(t) - L_i^r(t)C_i(t)-\sum_{j\in
      \mathbf{N}_i}K_{ij}^r(t)W_{ij})e_i \nonumber \\ &+&\!\!\sum_{j\in
      \mathbf{N}_i}K_{ij}^r(t)W_{ij}e_j +B(t)w-L_i^r(t)D_i(t)v_i  \nonumber \\ & 
-& \!\!\sum_{j\in
      \mathbf{N}_i}K_{ij}(t)^rH_{ij}v_{ij} 
      -F_if_i+F_i\varphi_i, \label{e} \\
e_i(0)&=&x_0-\xi_i. \nonumber
\end{eqnarray}
Noting that $f_i=\Upsilon_i\epsilon_i-\nu_i$, combine the dynamics of
the systems (\ref{e}) and (\ref{Om.sys.general}) into an augmented system
with $(e_i',\epsilon_i')'$ as a state vector:
 \begin{eqnarray}
    \dot{e}_i&=&(A(t) - L_i^r(t)C_i(t)-\sum_{j\in
      \mathbf{N}_i}K_{ij}^r(t)W_{ij})e_i \nonumber \\ &+&\sum_{j\in
      \mathbf{N}_i}K_{ij}^r(t)W_{ij}e_j -F_i\Upsilon_i\epsilon_i+F_i\varphi_i
    \nonumber \\ &+& 
B(t)w-L_i^r(t)D_i(t)v_i  
-\sum_{j\in
      \mathbf{N}_i}K_{ij}^r(t)H_{ij}v_{ij} 
      +F_i\nu_i,  \nonumber \\
\dot\epsilon_i &=& \Omega_i\epsilon_i+\Gamma_i \nu_i, \label{ext.e} \\
\hat f_i&=&\Upsilon_i\epsilon_i, \nonumber \\
&& e_i(0)=x_0-\xi_i, \qquad \epsilon_i(0)=0.  \nonumber
\end{eqnarray}
Observe that the system (\ref{ext.e}) at
node $i$ depends on the estimation errors at the neighboring nodes $j\in
\mathbf{N}_i$. 
Therefore, we propose a distributed observer of the form
(\ref{detector.general}) to estimate the state of the extended system
(\ref{ext.e}) and its outputs $\hat f_i$ from 
the outputs
(\ref{out.y}), (\ref{out.c}):
\begin{eqnarray}
    \dot{\hat{e}}_i&=&(A(t) - L_i^r(t)C_i(t)-\sum_{j\in
      \mathbf{N}_i}K_{ij}^r(t)W_{ij})\hat{e}_i \nonumber \\ 
& +& \sum_{j\in
      \mathbf{N}_i}K_{ij}^r(t)W_{ij}\hat e_j 
+\bar L_i^r(\zeta_i-C_i(t)\hat{e}_i)
    \nonumber \\ 
    &+& 
\sum_{j\in
      \mathbf{N}_i}\bar K_{ij}^r(t)(\zeta_{ij}-W_{ij}(\hat{e}_i-\hat{e}_j)), \nonumber \\
  \dot{\hat\epsilon}_i &=& 
\Omega_i \hat\epsilon_i +
    \check L_i^r(\zeta_i-C_i(t)\hat{e}_i) \nonumber \\
&+&\sum_{j\in
      \mathbf{N}_i}\check K_{ij}^r(t)(\zeta_{ij}-W_{ij}(\hat{e}_i-\hat
    e_j)), \label{ext.obs.nu.1.om.res}  \\ 
&&\hat{e}_i(0)=0, \quad \hat\epsilon_i(0)=0. 
\nonumber
\end{eqnarray}
The outputs of this observer 
\begin{equation}
  \label{varphi}
  \varphi_i= \Upsilon_i\hat{\epsilon}_i,   
\end{equation}
will
be constructed so that each output signal $\varphi_i$ approximates the 
attack input $f_i$ at the corresponding node $i$. This will allow it to be
used for feedback to compensate the attack as well as for signalling the
biasing attack. Note that the innovation signals (\ref{out.y}),
(\ref{out.c}) can be written as
\begin{eqnarray}
  \zeta_i&=&C_i(t)e_i + D_iv_i, 
                                              \label{out.y.1} \\
  \zeta_{ij}&=&-W_{ij}(e_j-e_i)+H_{ij}v_{ij}, \quad j\in \mathbf{N}_i, 
                                              \label{out.c.1}
\end{eqnarray}
and can be regarded as outputs of the interconnected large-scale
uncertain system comprised of systems (\ref{ext.e}). 


\subsection{The design algorithm}

To present the procedure for constructing a resilient observer of the form
(\ref{UP7.C.d.res}), let us consider the
error dynamics of the observer (\ref{ext.obs.nu.1.om.res}). Define 
$z_i=e_i-\hat e_i$, $\delta_i=\epsilon_i-\hat\epsilon_i$. Then it follows
from (\ref{ext.e}), (\ref{ext.obs.nu.1.om.res}) that 
\begin{eqnarray}
    \dot{z}_i&=&(A(t) - \hat L_i^r(t)C_i(t)-\sum_{j\in
      \mathbf{N}_i}\hat K_{ij}^r(t)W_{ij})z_i - F_i\Upsilon_i\delta_i
    \nonumber \\ && + \sum_{j\in
      \mathbf{N}_i}\hat K_{ij}^r(t)(W_{ij}z_j-H_{ij}v_{ij}) + Bw \nonumber \\ 
&& -\hat L_i^r(t)D_i(t)v_i 
      +F_i\nu_i, \nonumber \\
\dot \delta_i&=&\Omega_i \delta_i-\check L_i^r(t)C_i(t)z_i-\sum_{j\in
      \mathbf{N}_i}\check K_{ij}^r(t)W_{ij}z_i +\Gamma_i \nu_i\nonumber \\
&& + \sum_{j\in
      \mathbf{N}_i}\check K_{ij}^r(t)(W_{ij}z_j- H_{ij}v_{ij}) 
\label{ext.error.0} \label{ext.error} \\
&&  z_i(0)=x_0-\xi_i, \quad \delta_i(0)=0. \nonumber 
\end{eqnarray}
Here we used the notation 
\begin{eqnarray}
\hat L_i^r(t)=L_i^r(t)+\bar L_i^r(t), \quad 
\hat K_{ij}^r(t)=K_{ij}^r(t)+\bar K_{ij}^r(t).
\label{LK}
\end{eqnarray}
Although the equations describing the evolution of $z_i$ and $\delta_i$
look identical to the equations describing dynamics of the detector errors
in~\cite{DUSL1a,U10a}, the distinction 
lies in how $\hat L_i^r$, $\hat
K_{ij}^r$ are split to provide the gains for the state observer and the
attack detection filter at node $i$. 
Contrast to~\cite{DUSL1a,U10a}, in this paper the matrices
$L_i^r$, $K_{ij}^r$ are not considered to be given; they are determined
jointly with $\hat L_i^r$, $\hat K_{ij}^r$, $\check L_i^r$, $\check K_{ij}^r$
using the following procedure. 
 
\begin{enumerate}[1.]
\item
First, the coefficients $\hat L_i^r(t)$, $\hat K_{ij}^r(t)$, $\check
L_i^r(t)$, $\check K_{ij}^r(t)$ for each system (\ref{ext.error})  
are derived, to stabilize the uncertain interconnected system comprized
of the systems (\ref{ext.error}) in an
$\mathcal{L}_2$ sense. Then with these coefficients, one has
$\Upsilon_i(\epsilon_i-\hat\epsilon_i)\in\mathcal{L}_2[0,\infty)$.   
\item
The coefficients $L_i^r(t)$, $K_{ij}^r(t)$ for the controlled distributed
plant observer (\ref{UP7.C.d.res}) are computed in parallel with the
previous step. Since with the parameters derived in the previous
step, the signal
$f_i-\varphi_i=\Upsilon_i(\epsilon_i-\hat\epsilon_i)-\nu_i$ is  
$\mathcal{L}_2$ integrable for every admissible attack input $f_i$,  this
will be accomplished by treating $f_i-\varphi_i$ as a 
disturbance perturbing the error dynamics (\ref{e}). The coefficients
$L_i^r(t)$, 
$K_{ij}^r(t)$ are computed to attenuate these disturbances, along with
$w$, $v_i$, 
$v_{ij}$. Essentially, we redesign the original \emph{unbiased}
distributed plant observer (\ref{UP7.C.d.unbiased}) to make it is robust
against attack tracking errors which will arise as a result of applying the
attack cancelling control (\ref{contr}). 
 

\item
Finally, the remaining
coefficients $\bar L_i^r(t)$, $\bar K_{ij}^r(t)$ of the attack detector
(\ref{ext.obs.nu.1.om.res}) are obtained from (\ref{LK}), using the values 
$\hat L_i^r(t)$, $\hat K_{ij}^r(t)$ and $L_i^r(t)$, $K_{ij}^r(t)$ obtained
in the previous steps.   
\end{enumerate}



\subsubsection{Stabilization of the detector error dynamics
  (\ref{ext.error})}    
This step is identical to the corresponding step in~\cite{U10a}.  
Introduce the following notation:
\begin{eqnarray}
\mathbf{A}_i(t)&=&\left[\begin{array}{cc}
A(t) & -F_i\Upsilon_i\\
0 & \Omega_i
    \end{array}
  \right], \quad 
\mathbf{B}_i=\left[\begin{array}{cc}
B(t) & F_i\\
0 & \Gamma_i
    \end{array}
  \right], \nonumber \\
\mathbf{C}_i(t)&=&\left[\begin{array}{cc}
C_i(t) & 0 \\
W_{ij_1} & 0 \\
\vdots & \vdots \\
W_{ij_{q_i}} & 0
    \end{array}
  \right], \quad 
\mathbf{L}_i^r=\left[\begin{array}{cccccc}
\hat L_i^r & \hat K_{ij_1}^r & \ldots & \hat K_{ij_{q_i}}^r \\ 
\check L_i^r & \check K_{ij_1}^r & \ldots & \check K_{ij_{q_i}}^r
\end{array}\right], \quad \label{Ldef} 
\\
\mathbf{D_i}(t) &=& \left[\begin{array}{ccccccc} D_i(t) & 0 & \ldots & 0 & 0 & \ldots & 0  \\
                                 0   & H_{ij_1} & \ldots & 0 & Z_{ij_1}^{1/2}&
                                 \ldots & 0 \\ 
                                  \vdots & \vdots & \ddots & \vdots &
                                  \vdots & \ddots & \vdots \\
                                  0   & 0 & \ldots  & H_{ij_{q_i}} & 0 &
                                 \ldots & Z_{ij_{q_i}}^{1/2} 
                                      \end{array}\right]. 
\nonumber
\end{eqnarray}
It is assumed that $\mathbf{E}_i(t)=\mathbf{D_i}(t)\mathbf{D_i}'(t)>0$  for all $t$.  

Also, following~\cite{ZU1a}, introduce a collection of positive definite
$(n+n_{f_i})\times (n+n_{f_i})$ 
block-diagonal matrices 
$\mathbf{R}_i,\mathbf{X}_i$, $i=1\ldots,N$, partitioned as
\[
\mathbf{R}_i=\left[\begin{array}{cc}R_i & 0\\ 0 & \check R_i
  \end{array}\right],
\quad 
\mathbf{X}_i=\left[\begin{array}{cc}X_i & 0\\ 0 & \check X_i
  \end{array}\right],
  \] 
with $n\times n$ matrices $R_i$, $X_i$ and $n_{f_i}\times n_{f_i}$ matrices
$\check R_i$, $\check X_i$. Also, define the block matrix
$  \Phi=[\Phi_{ij}]_{i,j=1}^N,
$
\begin{eqnarray}
\Phi_{ij}=\begin{cases}\Delta_i, & i=j,\\
-W_{ij}'U_{ij}^{-1}W_{ij}, & i\neq j,~j\in\mathbf{N}_i,\\
0 & i\neq j,~j\not\in\mathbf{N}_i,
\end{cases}
\label{Phi.def} 
\end{eqnarray}
\begin{eqnarray}
U_{ij}=H_{ij}H_{ij}'+Z_{ij},\quad
\Delta_i=\sum_{j\in\mathbf{N}_i}W_{ij}'U_{ij}^{-1}Z_{ij}U_{ij}^{-1}W_{ij},
\label{Phi.def.1} 
\end{eqnarray}
$Z_{ij}$ $i=1,\ldots, N$, $j\in \mathbf{N}_i$ are square
$p_{ij}\times p_{ij}$ positive definite matrices. 
Also, let 
$R=\mathrm{diag}[R_1, \ldots, R_N]$, $\Delta=\mathrm{diag}[\Delta_1, \ldots, \Delta_N]$.
\begin{lemma}[cf.~\cite{U10a,ZU1a}]\label{T1}
Suppose there exists a constant $\gamma>0$ and 
symmetric matrices $R_i\ge 0$, $\check R_i\ge 0$, $Z_{ij}>0$,
$j\in \mathbf{N}_i$, $i=1, \ldots N$, such that  
\begin{enumerate}[(i)]
\item
the following linear matrix inequalities are satisfied
\begin{equation}
\label{LMI}
  R+\gamma^2(\Phi+\Phi'-\Delta) > 0, \quad \check
  R_i>\Upsilon_i'\Upsilon_i;
\end{equation}
\item
each differential Riccati equation  
\begin{eqnarray}\label{Riccati} 
\dot{\mathbf{Y}}_i &=& \mathbf{A}_i\mathbf{Y}_i+\mathbf{Y}_i\mathbf{A}_i'
\nonumber \\
&& - \mathbf{Y}_i(\mathbf{C}_i'\mathbf{E}_i^{-1}\mathbf{C}_i
-\frac{1}{\gamma^2}\mathbf{R}_i)\mathbf{Y}_i
+\mathbf{B}_i\mathbf{B}_i', \qquad  \\
\mathbf{Y}_i(0)&=&\mathbf{X}_i^{-1}, \nonumber  
\end{eqnarray}
has a positive definite symmetric bounded solution $\mathbf{Y}_i(t)$ on the
interval $[0,\infty)$, i.e., for all $t\ge 0$,
$\alpha_1I<\mathbf{Y}_i(t)=\mathbf{Y}_i'(t)<\alpha_2I$, for some $\alpha_{1,2}>0$.  
\end{enumerate}
Then the network of systems (\ref{ext.error}) 
 with the coefficients $\hat L_i^r$, $\hat K_{ij}^r$, $\check L_i^r$,
 $\check K_{ij}^r$, obtained by partitioning the matrices
\begin{eqnarray}
  \label{L}
  \mathbf{L}_i^r(t)=\mathbf{Y}_i(t)\mathbf{C}_i(t)'\mathbf{E}_i^{-1}(t).
\end{eqnarray}
according to (\ref{Ldef}), guarantees that the noise- and attack-free network
is exponentially stable, and in the presence of disturbances or an attack
it holds that  
\begin{eqnarray}
\lefteqn{\sum_{i=1}^N\int_0^{\infty}\|\hat f_i-\varphi_i\|^2dt\le \gamma^2 \sum_{i=1}^N\bigg((x_0-\xi_i)'X_i^{-1}(x_0-\xi_i)} &&
\nonumber \\
&&+ \int_0^\infty\!\!\big(\|w\|^2+\|v_i\|^2+\|\nu_i\|^2+\sum_{j\in
    \mathbf{N}_{ij}}\|v_{ij}\|^2\big)dt\bigg). \quad 
\label{z.Hinf}
\end{eqnarray}
\end{lemma}

The proof of Lemma~\ref{T1} is omitted for brevity, it uses a 
completion of squares argument to establish that
$V=\sum_{i=1}^N[z_i'~\delta_i']\mathbf{Y}_i^{-1}(t)\left[\begin{array}{c}z_i\\
    \delta_i
  \end{array}\right]$ is a Lyapunov function
for the large-scale system comprised of systems (\ref{ext.error}). Also, it trivially
follows from (\ref{z.Hinf}) that each signal 
\begin{equation}
  \label{eta}
\eta_{ij}=-W_{ij}z_j, \quad j\in\mathbf{N}_i, 
\end{equation}
is $\mathcal{L}_2$-integrable; each such signal $\eta_{ij}$ connects 
the system (\ref{ext.error}) at node $i$ with the analogous system at 
node $j$, $j\in\mathbf{N}_i$. It then follows from Lemma~\ref{T1},
condition (ii), that each detector ensures the following \emph{decentralized
disturbance attenuation} performance:
\begin{eqnarray}
\lefteqn{\int_0^{\infty}(\|z_i\|^2_{R_i}+\|\delta_i\|^2_{\check R_i})dt} &&
\nonumber \\
&&\le \gamma^2 \bigg((x_0-\xi_i)'X_i^{-1}(x_0-\xi_i)
\nonumber \\
&&+ \int_0^\infty\!\!\big(\|w\|^2+\|v_i\|^2+\sum_{j\in
    \mathbf{N}_{ij}}(\|\eta_{ij}\|_{Z_{ij}^{-1}}^2+\|v_{ij}\|^2\big)dt\bigg). \quad 
\label{z.Hinf.local}
\end{eqnarray}
This explains the role of the matrices $Z_{ij}$ as weights on the
contribution of the interconnection signals $\eta_{ij}$ in the individual
performance of each detector component~(\ref{ext.obs.nu.1.om.res}).

According to Lemma~\ref{T1}, each node computes the matrix
$\mathbf{L}_i^r$ and its components $\hat L_i^r$, $\hat K_{ij}^r(t)$, $\check
L_i^r$, $\check K_{ij}^r(t)$ independently from other nodes. For this, 
the respective Riccati differential equation (\ref{Riccati}) must be solved
on-line; this allows the matrix $\mathbf{L}_i^r$ to be computed and
partitioned according to (\ref{Ldef}) in real time. Unlike~\cite{DUSL1a},
the nodes do not need to communicate to solve these Riccati
equations. To setup these equations, the matrices $R_i$ must be
determined first from the LMIs~(\ref{LMI}). Even though this step must be
performed centrally, it does not require the knowledge of the parameters of the
system observed; only the matrices $W_{ij}$ and $H_{ij}$
are required which characterize the communication network. 
Compared with~\cite{DUSL1a}, this reduces
substantially the amount 
of information that the node must agree upon in advance. As long as the
matrices $W_{ij}$, $H_{ij}$ and $Z_{ij}$ and the disturbance attenuation
parameter $\gamma^2$ do not change, the same matrices $R_i$ and $X_i$ can
be utilized even when the plant changes substantially. In the case of such
an event, each node must only update its Riccati equation
(\ref{Riccati}); it can do so without communicating with its neighbours.   


\subsubsection{Design of the resilient distributed plant observer (\ref{UP7.C.d.res})}    

Now that we able to guarantee that 
$f_i-\varphi_i=(\hat f_i-\varphi_i)-\nu_i\in \mathcal{L}_2$, 
the large-scale system comprised of 
the error dynamics (\ref{e}) of the observer (\ref{UP7.C.d.res}),
(\ref{contr}) can be stabilized in an $H_\infty$ sense,
while attenuating this disturbance. 
The coefficients $L_i^r$, $K_{ij}^r$ which accomplish this
task can
be computed in parallel with the coefficients of the attack detector, using the
same approach based on the results of~\cite{ZU1a}. To present this step of
our algorithm, introduce the notation 
\begin{eqnarray}
 \mathbf{B}_{1,i}&=&\left[\begin{array}{cc}
B(t) & F_i
    \end{array}
  \right], \quad
\mathbf{C}_{1,i}(t)=\left[\begin{array}{cccc}
C_i'(t)  &
W_{ij_1}' &
\ldots  &
W_{ij_{q_i}}'
    \end{array}
  \right]', \nonumber \\ 
\mathbf{L}_{1,i}&=&\left[\begin{array}{cccccc}
L_i^r & K_{ij_1}^r & \ldots & K_{ij_{q_i}}^r
\end{array}\right].
\label{Ldef.prime}
\end{eqnarray}

\begin{lemma}[see~\cite{ZU1a}]\label{T2}
Suppose there exists a constant $\bar\gamma>0$ and 
symmetric matrices $\bar R_i\ge 0$, $\bar X_i> 0$, $Z_{ij}>0$, $j\in
\mathbf{N}_i$, $i=1, \ldots N$, such that  
\begin{enumerate}[(i)]
\item
the following linear matrix inequality is satisfied
\begin{equation}
\label{LMI.1}
  \bar R+\bar\gamma^2(\Phi+\Phi'-\Delta) > P,
\end{equation}
where $\bar R=\mathrm{diag}[\bar R_1,\ldots,\bar R_N]$, and 
$\Phi$, $\Delta$ are the matrices defined in (\ref{Phi.def}),
(\ref{Phi.def.1}) which are the same as in
Lemma~\ref{T1}\footnote{Performance tuning of
  the algorithm may require one to choose different matrices $Z_{ij}$ in this step of the algorithm. In this case, the matrices $\Phi$,
  $\Delta$, and $\mathbf{E}_i$ will also need to be updated, and will not
  be the same as in Lemma~\ref{T1}. However, this does not have any effect
  on the statement of Lemma~\ref{T2}.};  
\item
each differential Riccati equation  
\begin{eqnarray}\label{Riccati.1} 
\dot{Y}_i &=& AY_i+Y_iA'
\nonumber \\
&& - Y_i(\mathbf{C}_{1,i}'\mathbf{E}_i^{-1}\mathbf{C}_{1,i}
-\frac{1}{\bar\gamma^2}\bar R_i)Y_i
+\mathbf{B}_{1,i}\mathbf{B}_{1,i}', \qquad  \\
&&Y_i(0)=\bar X_i^{-1}, \nonumber  
\end{eqnarray}
has a positive definite symmetric bounded solution $Y_i(t)$ on the
interval $[0,\infty)$, i.e., for all $t\ge 0$,
$\bar\alpha_1I<Y_i(t)=Y_i'(t)<\bar\alpha_2I$, for some $\bar\alpha_{1,2}>0$.  
\end{enumerate}
Then the network of systems (\ref{e}),  
 with the coefficients $L_i^r$, $K_{ij}^r$, obtained by partitioning the
 matrices 
\begin{eqnarray}
  \label{L1}
  \mathbf{L}_{1,i}^r(t)=Y_i(t)\mathbf{C}_{1,i}(t)'\mathbf{E}_i^{-1}(t).
\end{eqnarray}
according to (\ref{Ldef.prime}), guarantees that the noise- and attack-free
network of error dynamics (\ref{e}) is exponentially stable, and in the presence
of disturbances or an attack it holds that  
\begin{eqnarray}
\lefteqn{\int_0^{\infty}\mathbf{e}'P\mathbf{e}dt\le \bar\gamma^2
  \sum_{i=1}^N\bigg((x_0-\xi_i)'X_i^{-1}(x_0-\xi_i)} && 
\nonumber \\
&&+ \int_0^\infty\!\!\big(\|w\|^2+\|v_i\|^2+\|\varphi_i-f_i\|^2+\sum_{j\in
    \mathbf{N}_{ij}}\|v_{ij}\|^2\big)dt\bigg). \quad 
\label{z.Hinf.prime}
\end{eqnarray}
\end{lemma}

The proof of the lemma is analogous to the proof of the corresponding
result in~\cite{ZU1a}. 

\subsubsection{The main result}
    
The main result of this paper follows from the properties of the observer
errors (\ref{e}) and the properties of the errors of the 
attack detection filters (\ref{ext.obs.nu.1.om.res}).

\begin{theorem}\label{main}
Suppose the conditions of Lemmas~\ref{T1} and~\ref{T2} hold. Let the
coefficients $\check L_i^r$, $\check K_{ij}^r$ of the detectors
(\ref{ext.obs.nu.1.om.res}) be obtained as described in Lemma~\ref{T1}, and
let the coefficients $\bar L_i^r$, $\bar K_{ij}^r$ be obtained using the
matrices $\hat L_i^r$, $\hat K_{ij}^r$, $L_i^r$, $K_{ij}^r$ from
Lemmas~\ref{T1} and~\ref{T2},   
as
\begin{eqnarray}
  \label{barK}
  \bar L_i^r=\hat L_i^r-L_i^r, \quad \bar K_{ij}^r=\hat K_{ij}^r-K_{ij}^r. 
\end{eqnarray}
Then, the network of state observers (\ref{UP7.C.d.res}), augmented with
the network of attack detectors (\ref{ext.obs.nu.1.om.res}) produces state
estimates $\hat x_i$ which have the following convergence properties. 
\begin{enumerate}[(i)]
\item
In the absence of disturbances and biasing attacks, $\|x-\hat x_i\|\to 0$
exponentially as $t\to\infty$. 
\item
When the plant and/or the network is subject to $\mathcal{L}_2$-integrable
disturbances and/or admissible biasing attacks, the estimates $\hat x_i$
converge to $x$ in the $\mathcal{L}_2$ sense,
and the resilient performance of this observer network is characterized by the
condition
\begin{eqnarray}
\lefteqn{\int_0^{\infty}\mathbf{e}'P\mathbf{e}dt}
&& 
\nonumber \\
&&\le \bar\gamma^2
  \sum_{i=1}^N\bigg((x_0-\xi_i)'(\bar X_i^{-1}+2\gamma^2 X_i^{-1})(x_0-\xi_i) \nonumber \\
&&+ (1 + 2\gamma^2)\int_0^\infty\!\!\big(\|w\|^2+\|v_i\|^2+\sum_{j\in
    \mathbf{N}_{ij}}\|v_{ij}\|^2\big)dt\bigg) \nonumber \\
&&+ 2\bar\gamma^2 (1+\gamma^2) \sum_{i=1}^N
\int_0^\infty\|\nu_i\|^2dt.
\label{z.Hinf.final}
\end{eqnarray}
Also, the outputs $\varphi_i$ of the
distributed attack detector network (\ref{ext.obs.nu.1.om.res}) track the
attack inputs in the $\mathcal{L}_2$ sense.
\end{enumerate}
\end{theorem}

\section{Conclusion} \label{sec:conclusion}
We have proposed a novel class of distributed
observers for robust estimation of a linear plant, which are resilient 
to biasing misappropriation attacks. The observers involve feedback from an
additional network of attack detection filters, which can also signal the
attack. The design method is based on the methodology of distributed and
decentralized $H_\infty$ filtering which is combined with a decoupling
technique to obtain observers which attenuate benign disturbances, while
sensing and compensating biasing inputs.


\newcommand{\noopsort}[1]{} \newcommand{\printfirst}[2]{#1}
  \newcommand{\singleletter}[1]{#1} \newcommand{\switchargs}[2]{#2#1}

\end{document}